\documentclass{vgtc}                          

\graphicspath{{figures/}{pictures/}{images/}{./}} 

\usepackage{times}                     

\usepackage{tabu}                      
\usepackage{booktabs}                  
\usepackage{lipsum}                    
\usepackage{mwe}                       

\usepackage{mathptmx}                  

\onlineid{0}

\vgtccategory{Research}

\vgtcinsertpkg

\title{A Part-to-Whole Circular Cell Explorer}

\author{Siyuan Zhao\thanks{e-mail: szhao69@uic.edu}\\ %
        \scriptsize University of Illinois Chicago %
\and G. Elisabeta Marai\thanks{e-mail: gmarai@uic.edu}\\ %
     \scriptsize University of Illinois Chicago %
}

\teaser{
  \centering
  \includegraphics[width=0.8\linewidth]{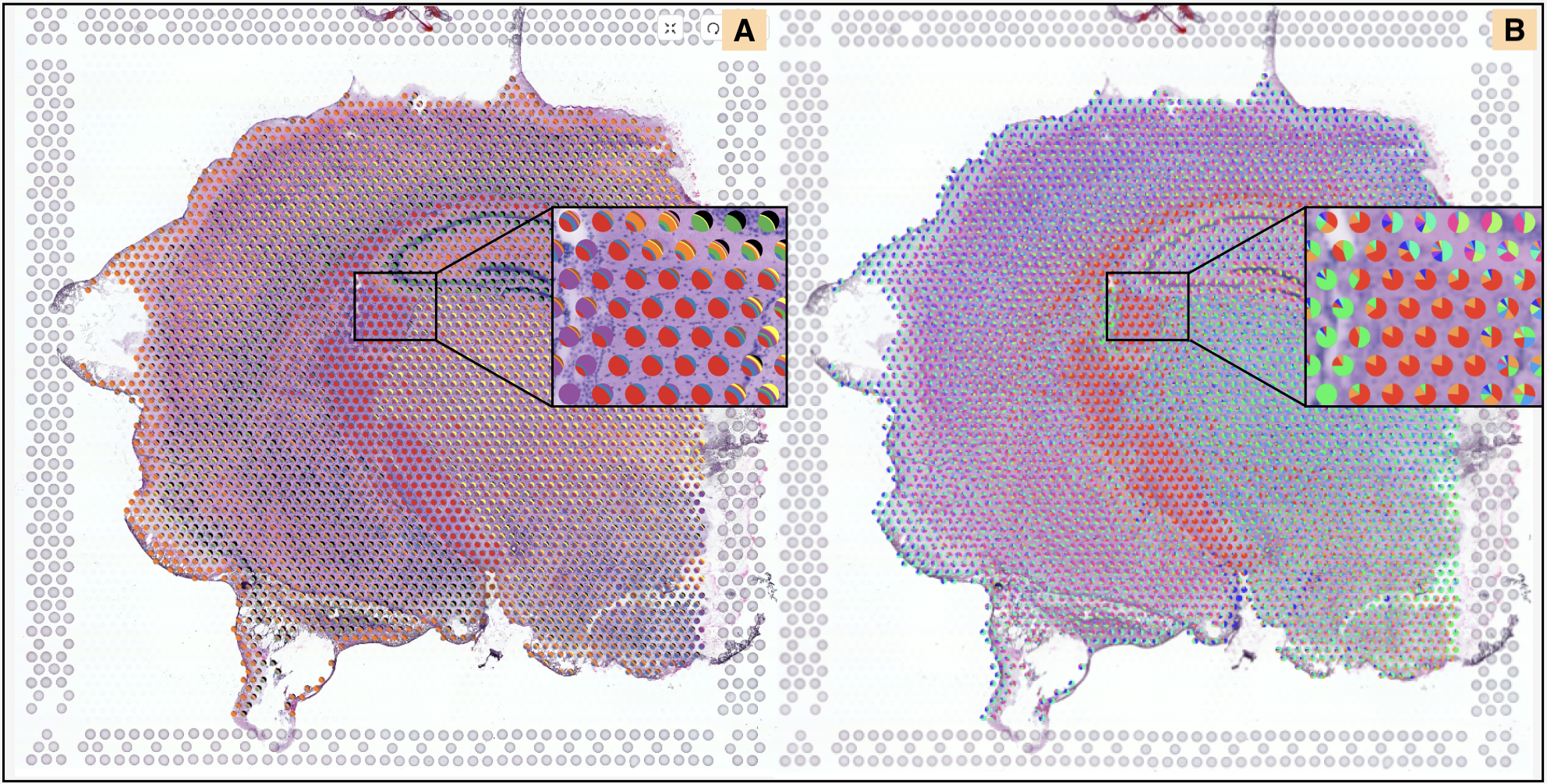}
  \caption{Part-to-Whole circular slice encodings of the gene type distribution in a cell region of interest, with corrected color scheme (A) versus pie-chart encodings of the gene distribution in the same region of interest, with the original color scheme (B). Introduced by Kosara in 2019 [2], the circular slice encoding has a compact design with small footprint, comparable error to linear barcharts, lower absolute error than the pie chart, and is faster to read than pie charts as well. Note how in the revised illustration (A) the inset distributions are easier to read, and the main structures in the image are still visible, while the interplay between the dominant cell types (e.g., area below the inset) is far easier to see than in (B). 
}
  \label{fig:comparation-kosara-piechart}
}

\abstract{
    Spatial transcriptomics methods capture cellular measurements such as gene expression and cell types at specific locations in a cell, helping provide a localized picture of tissue health. Traditional visualization techniques superimpose the tissue image with pie charts for the cell distribution. We design an interactive visual analysis system that addresses perceptual problems in the state of the art, while adding filtering, drilling, and clustering analysis capabilities. Our approach can help researchers gain deeper insights into the molecular mechanisms underlying complex biological processes within tissues.
} 

\begin{document}


\firstsection{Introduction}

\maketitle

Combining tissue images and cell type membership ratio results with spatial transcriptomics can reveal spatial patterns characteristics of health and disease. While domain experts often superimpose cell type pie charts over H\&E stained tissue images, these encodings have perceptual limitations~\cite{munzner2014visualization}. We design an interactive visual analysis system that leverages a performant circular slice encoding and a customizable color palette, in addition to three coordinated views to holistically visualize gene expression and cell type under the given tissue. Our approach retains the crucial spatial context, while resolving the issue of pie chart perception. The resulting system helps researchers navigate the intricate dimensions of gene expression and cell types.

\section{Related Work}
In spatial transcriptomics visualization, the most common visual representation aims to identify cell types and cell states by visualizing as colormaps gene expression results in the areas of interest. Several packages, including Giotto~\cite{Dries2021}, Scanpy~\cite{Wolf2018}, Seurat~\cite{Hao2021}, and Squidpy~\cite{Palla2022} offer functionalities to plot spatial transcriptomic data by visualizing the spatial gene expression as colormaps. However, a notable limitation of these tools is that they do not explicitly address the exploration of cell type composition within each spot. On the other hand, Song et al.~\cite{Song2021} use a Graph Convolutional Network (GCN)~\cite{Kipf2016} to display the cellular compositions as a pie chart at each spot. However, the pie chart encoding has perceptual limitations~\cite{munzner2014visualization}, as human perception of radial layouts is typically inferior to linear layouts. Our work also integrates visualization for both gene expression and cellular compositions, while leveraging a perceptually-improved design and additional interactive statistical views for analyzing the cell type distributions.  

\section{Redesign}
\begin{figure*}[ht]
\centering
\includegraphics[width=0.8\linewidth]{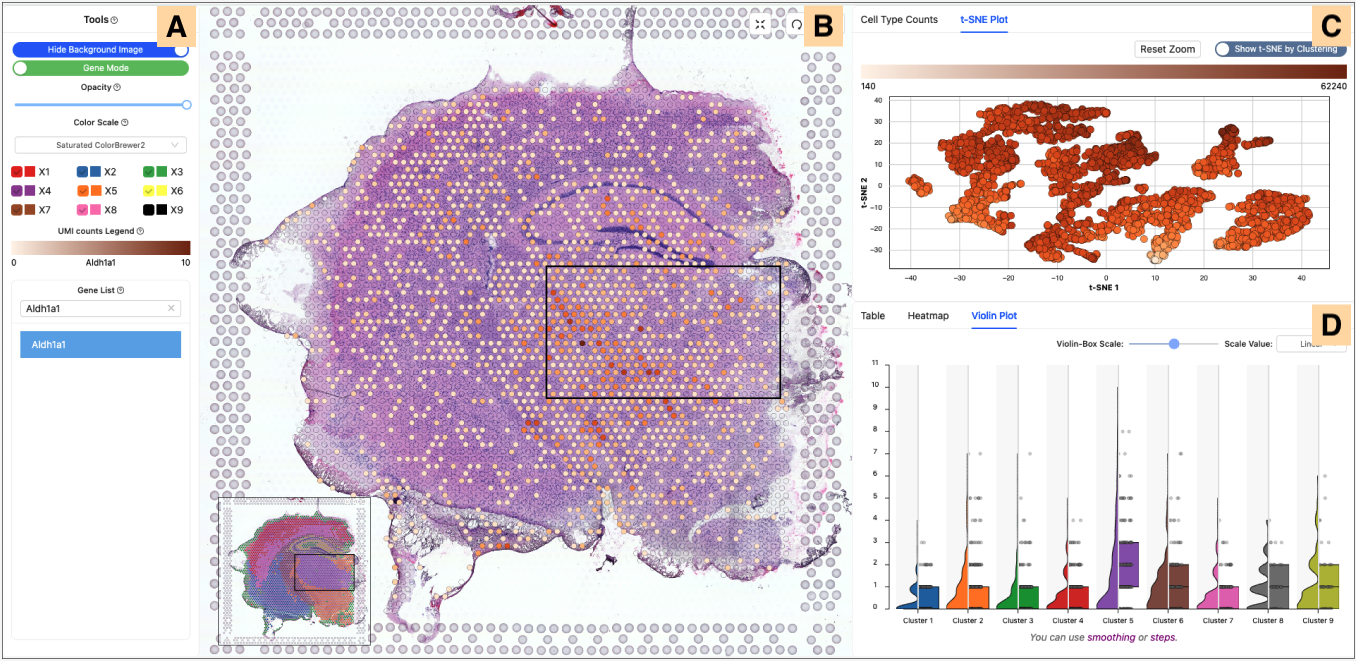}
\caption{Interactive web-based tool leveraging the redesigned encoding: (A) a control panel allows changing the opacity of the tissue image and circular slice encodings, as well as selecting a subset of genes to view and analyze; (B) tissue image view, showing here the result of t-SNE~\cite{Laurens2008} clustering and gene expression, currently showing gene expression of regional THAL\_lat astrocytes markers~\cite{kleshchevnikov2022}; (C) projection view; (D) hybrid violin plot for the distribution of each cluster, which can also be explored in tabular or heatmap form.}
\label{fig:gene-expression-astrocyte-marker}
\end{figure*}

We created an interactive web-based tool with three coordinated views, implemented in D3 with Flask (Fig.~\ref{fig:gene-expression-astrocyte-marker}.

\textbf{Tissue image view}. We leverage a part-to-whole circular slice encoding~\cite{Kosara2019} (Fig.~\ref{fig:comparation-kosara-piechart}, in which part-to-whole values are represented by sliding circles of the same size over a base circle. This encoding has been shown to outperform stacked bar charts~\cite{Kosara2019}. This redesign preserves the compact design with a small footprint, while providing good readability in terms of error and time. We also replaced the original saturated rainbow color scheme with a qualitative scheme from colorbrewer2~\cite{Harrower2013}, which allows increased legibility of the background image layer, although we also support customization of the color map. The opacity of the two layers can be interactively adjusted.

\begin{figure}[ht]
\centering
\includegraphics[width=0.8\linewidth]{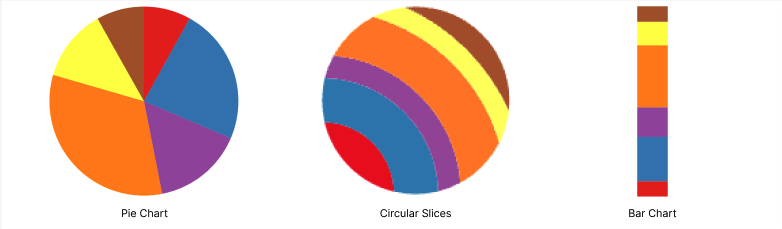}
\caption{Pie chart, circular slices, and stacked bar chart}
\label{fig:kosara-comparation}
\end{figure}

\textbf{Projection View}. This view complements the spatial layout by supporting analysis of the similarity of gene expression profiles in the context of primary cell-type composition. We use t-SNE~\cite{Laurens2008} to convert the gene expression matrix (cells x genes) to a 2-dimensional projection. This type of projection can provide insight into the molecular mechanisms underlying complex biological processes within tissues (Fig.~\ref{fig:gene-expression-astrocyte-marker} inset).

 \textbf{Gene View}. This view allows users to search for specific genes and visualize their expression based on the selected UMI calculation method, revealing the distribution of expression across the tissue across different calculation methods (linear, log2, logNorm), similar to Seurat~\cite{Hao2021} and Scanpy~\cite{Wolf2018}). Typically, regions with higher UMI counts suggest greater RNA content compared to those with lower counts. The view supports analysis of gene expression through three integrated components. The Up-Regulated Genes Table captures significant gene expression changes within each cluster. The Heatmap shows L2FC values along with the distribution and magnitude of gene expression changes. The Violin Plot (Fig.~\ref{fig:gene-expression-astrocyte-marker}) shows the localization of gene expression in regional THAL\_lat astrocytes, marked by Aldh1a1 expression. THAL\_lat astrocytes, which are primarily found in the ventrolateral thalamic nucleus, show significant expression of Aldh1a1 in Cluster 5 (the purple area in the t-SNE clustering inset). 


\section{Discussion and Conclusion}
Our preliminary results (Fig.~\ref{fig:comparation-kosara-piechart} and Fig.~\ref{fig:gene-expression-astrocyte-marker} show this interactive visual computing system addresses a challenging issue in spatial transgenics: presenting cellular information while maintaining spatial context. 

Our solution is not without limitations. Our use of colormaps is perceptually limiting, as humans can distinguish among about only 6 or 7 colors, which inherently results with larger datasets in spurious visual similarity among similar colors. Likewise, zooming on the tissue image can result in a loss of global context. 

Nevertheless, our solution offers detailed and comprehensive visualizations of gene expression for each spatial spot, and improves upon the state of the art. This approach can empower domain experts to gain insights and extract useful information.

\acknowledgments{
The authors are supported by NIH (UG3-TR004501 and NCI-R01-CA258827), and NSF (CNS-2320261). 
}

\bibliographystyle{abbrv-doi}

\bibliography{template}

\begin{thebibliography}{10}

\bibitem{Dries2021}
R.~Dries, Q.~Zhu, R.~Dong, C.-H.~L. Eng, et~al.
\newblock Giotto: a toolbox for integrative analysis and visualization of spatial expression data.
\newblock {\em Genome biology}, 22:1--31, 2021.

\bibitem{Hao2021}
Y.~Hao, S.~Hao, E.~Andersen-Nissen, W.~Mauck, et~al.
\newblock Integrated analysis of multimodal single-cell data.
\newblock {\em Cell}, 184(13):3573--3587, 2021.

\bibitem{Harrower2013}
M.~Harrower and C.~Brewer.
\newblock Colorbrewer. org: an online tool for selecting colour schemes for maps.
\newblock {\em Cartogr. J.}, 40(1):27--37, 2003.

\bibitem{Kipf2016}
K.~Kipf and M.~Welling.
\newblock Semi-supervised classification with graph convolutional networks.
\newblock In {\em Int. Conf. Lear. Repres.}, 2017.

\bibitem{kleshchevnikov2022}
V.~Kleshchevnikov, A.~Shmatko, E.~Dann, A.~Aivazidis, et~al.
\newblock Cell2location maps fine-grained cell types in spatial transcriptomics.
\newblock {\em Nat. Biotechnol.}, 40(5):661--671, 2022.

\bibitem{Kosara2019}
R.~Kosara.
\newblock The impact of distribution and chart type on part-to-whole comparisons.
\newblock In {\em Short Pap. Proc. Eurographics/IEEE VGTC Symp. Vis. (EuroVis)}, 2019. doi: {{%
10\hspace{.1pt}\discretionary{.}{%
}{.}\hspace{.4pt}2312\discretionary{/}{%
}{/}evs20191162}}


\bibitem{munzner2014visualization}
T.~Munzner.
\newblock {\em Visualization analysis and design}.
\newblock CRC press, 2014.

\bibitem{Palla2022}
G.~Palla, H.~Spitzer, M.~Klein, D.~Fischer, et~al.
\newblock Squidpy: a scalable framework for spatial omics analysis.
\newblock {\em Nat. methods}, 19(2):171--178, 2022.

\bibitem{Song2021}
Q.~Song and J.~Su.
\newblock Dstg: deconvoluting spatial transcriptomics data through graph-based artificial intelligence.
\newblock {\em Brief. Bioinform.}, 22(5):bbaa414, 2021.

\bibitem{Laurens2008}
L.~{v. d. M.} and G.~{E. H.}
\newblock Visualizing data using t-sne.
\newblock {\em J. Mach. Learn. Res.}, 9:2579--2605, 2008.

\bibitem{Wolf2018}
F.~Wolf, P.~Angerer, and F.~Theis.
\newblock Scanpy: large-scale single-cell gene expression data analysis.
\newblock {\em Genome biol.}, 19:1--5, 2018.

\end{thebibliography}

\end{document}